\documentstyle[aps,prl,epsf]{revtex}

\begin{document}
\draft
\twocolumn[\hsize\textwidth\columnwidth\hsize\csname
@twocolumnfalse\endcsname
\title{A Kondo impurity in a disordered metal: Anderson's theorem
revisited}
\author{Sudip Chakravarty and Chetan Nayak}
\address{Department of Physics and Astronomy, University of
California
Los Angeles\\
Los Angeles, CA 90095-1547}
\date{\today}
\maketitle
\begin{abstract}
We consider a local moment which is coupled by
a non-random Kondo $J$ to a band of conduction electrons
in a random potential. We prove an analog of
Anderson's theorem in a large-$N$ limit of this model.
The theorem states that when the disorder is weak,
the disorder-averaged low-temperature thermodynamics
is independent of the strength of the disorder;
remarkably, it further states that fluctuation effects in the long-time
limit
are {\it independent even of the realization
of the disorder}. We discuss the
relationship of this theorem to theoretical and
experimental studies of similar problems.
\end{abstract}
]
There are not many known  examples for which the quantum mechanical
ground state is protected from
perturbations due to disorder. There are equally
few examples for which
the effect of the said disorder can be understood
in a simple, but precise manner. A
$s$-wave superconductor in the
presence of nonmagnetic impurities
furnishes an example of both, and its
resilience with respect to
disorder can be understood from
Anderson's theorem\cite{Anderson1}. The
theorem can be stated as the
independence of the order parameter, and
{\em inter alia} the transition temperature, of the
randomness, unless the single particle electronic states are themselves
localized.

As remarked by Markowitz and Kadanoff\cite{Markowitz}, the theorem
avoids a number of
pitfalls, each of which could give rise to rather strong effects of
disorder on the superconducting
transition temperature. The key insight is that the superconducting
state is a broken
symmetry state built from time reversed pairs of single particle states.
Being a broken symmetry
state, its order parameter should be calculable from a mean field
equation away from regimes in which
fluctuations dominate, and this equation remains unchanged in the
presence of potential scattering due
to impurities. The point is that no matter how complicated these single
particle states are in the disorder potential, the order parameter
depends only on the average density
of these states, and therefore it remains
unchanged. The theorem could fail if
the fluctuations of the order parameter are
strong, and the assumed mean field
theory no longer holds.

In this Letter, we construct an analog of Anderson's theorem in a
problem, which appears to be even
more complex, because there is no broken symmetry. This is
the ``Kondo" problem, or a class of Kondo problems of a magnetic moment
in a disordered metal. A
perturbative  approach\cite{Martin} leads to a
great many difficulties
arising from the quenched randomness
combined with the marginality of the perturbation theory in the Kondo
coupling constant $J$. But,
paradoxically, the low temperature
strong coupling behavior, which is usually  difficult from the
theoretical
perspective, appears to be  exceedingly simple. The key here is the
nature of the strong coupling fixed
point, which protects the system from disorder despite
the absence of broken symmetry. We access this strong coupling
fixed point in a controlled way with a large-$N$ limit of the Kondo
problem, but we believe that the theorem holds
more generally. In this paper we shall not treat the
weak-coupling behavior, nor will
we discuss the detailed crossover from  weak to strong coupling. The
reason is that there is very
little to protect us from the complexities of the weak coupling
perturbation theories of this problem. 

We  also address the relationship of our work to
recent perplexing experimental studies and show that these experiments
can possibly be understood on the basis of our theorem. 

Consider the orbitally degenerate  Coqblin-Schrieffer
Hamiltonian\cite{Coqblin} in the
presence of random potential scattering
as the generic Hamiltonian,
which contains all the important aspects of the problem.
The Hamiltonian is
\begin{eqnarray}
H=\sum_{kM} \varepsilon_k
c^{\dagger}_{kM}c_{kM}&+&\sum_{kk'M} V_{kk'}
c^{\dagger}_{k'M}c_{kM}\nonumber \\
&-&J\sum_{kk'MM'}c^{\dagger}_{k'M'}f^{\dagger}_{M}f_{M'}c_{kM}.
\end{eqnarray}
The  operator  $c^{\dagger}_{kM}$ creates a conduction electron of wave
number $k$, total
angular momentum $j$ and a $z$-component of the angular momentum
$j_z=M$.  We have assumed that the
origin is at the location of the magnetic impurity, so that
$\sum_k c^{\dagger}_{kM}$
defines the Wannier operator
$\psi^{\dagger}_M(0)$ at the origin. The operator
$f^{\dagger}_{M}$ creates an electron on a magnetic impurity embedded
within the conduction electrons with
the same $j$ and the $z$-component $M$.
The magnetic quantum number can take any value $\Delta M=M-M'$ and
is not restricted to $\pm 1$,
or 0 as in a spin-1/2 model. The total number of states is
$N=(2j+1)$. The random potential
$V_{kk'}$ describes the disordered conduction electrons, where
$V(r)=\sum_iU(r-R_i)$, and
$U(r-R)$ is the potential due to a single impurity at $R$.

We shall treat this model
in the limit that the degeneracy $N$ of the
orbitals tends to
$\infty$\cite{largeN}, but the presence of
quenched disorder must be
taken into account. The method is
particularly suitable for accessing the
low-temperature strong coupling
regime, not available to
perturbative methods. To define a
smooth large-$N$ limit, we impose the
condition that
$\sum_{M}f^{\dagger}_{M}f_{M}=N/2$\cite{largeN,Coleman} instead of
$\sum_{M}f^{\dagger}_{M}f_{M}=1$\cite{Read,Read2},
allowing the use of the small parameter $1/N$ as a mathematical device.

Let us make a change of basis to the exact eigenstates of the electrons
in the presence of {\em a given
realization} of the random potential,
but without the magnetic impurity, where
\begin{equation}
c_{kM}=\sum_{\alpha} \langle k|\alpha\rangle  c_{\alpha M}
\end{equation}
The transformed Hamiltonian is thus
\begin{equation}
H=\sum_{\alpha M} \varepsilon_\alpha
c^{\dagger}_{\alpha M}c_{\alpha M}
-J\sum_{\alpha
\alpha'MM'}a^*_{\alpha'}a_{\alpha}c^{\dagger}_{\alpha'M'}
f^{\dagger}_{M}f_{M'}c_{\alpha M},
\end{equation}
where the amplitude $a_\alpha=\sum_k \langle k|\alpha\rangle$; inserting
a complete set of position kets, it is
easily seen that $a_\alpha$ is nothing but the exact single particle
{\em wave function}  at the
origin, $\phi_\alpha(0)$.

As a first step, we calculate the partition function $Z[V]$ in a large-$N$ expansion for a
given realization of the random
potential, where
\begin{equation}
Z[V]=\int {\cal D}c{\cal D}c^*{\cal D}f{\cal
D}f^*e^{-\int_0^{\beta}d\tau
\cal{L}(\tau)}\prod_{\tau}\delta(n_f(\tau)-N/2).
\end{equation}
The Euclidean Lagrangian in terms of the Grassman variables $c, c^*, f,
f^*$ is given by
\begin{eqnarray}
{\cal{L}}(\tau)&=&\sum_{\alpha
M}c^*_{\alpha
M}\left(\frac{\partial}{\partial\tau}+\varepsilon_\alpha\right)c_{\alpha
M}+\sum_M
f^*_M\frac{\partial}{\partial\tau}f_M\nonumber \\
&-&J\sum_{\alpha \alpha'MM'}a^*_{\alpha'}a_\alpha
c^{*}_{\alpha'M'}f^{*}_{M}f_{M'}c_{\alpha M}
\end{eqnarray}
The $\delta$-function constraint on the occupation of the $f$-level can
be resolved by introducing another
functional integral involving a field $\lambda(\tau)$, and the fermionic
degrees of freedom can be integrated
out by introducing a complex Hubbard-Stratonovich field $\sigma(\tau)$,
which plays the role of the hybridization in a $U=0$
Anderson model:
\begin{equation}
\left\langle\sigma\right\rangle =
\left\langle{\sum_{\alpha,M}} {a_\alpha}{f_M^*}\,{c_{\alpha M}}
\right\rangle
\end{equation}
The result is\cite{largeN,Read,Coleman2}
\begin{equation}
Z[V]=Z_0[V]\int {\cal D \sigma}{\cal D \sigma^*}{\cal D
\lambda}\exp(-S_{\rm eff}),
\end{equation}
where $Z_0[V]$ is the partition function
of the disordered electronic system
without the magnetic impurity and
\begin{eqnarray}
S_{\rm eff}&=&-N{\rm
Tr}\ln\left(\frac{\partial}{\partial\tau}+i\lambda+\sigma^*
G^0\sigma\right)\nonumber \\&+&\int_0^{\beta}
d\tau\left(\frac{N}{J_0}|\sigma|^2-i\lambda\right).
\end{eqnarray}
We have defined $J_0=-NJ=N|J|$ for the antiferromagnetic problem of our
present interest. The local Green
function, $G_0(\tau)$, at the impurity site is
\begin{equation}
G^0(\tau)=-\sum_\alpha|a_\alpha|^2\left(
\frac{\partial}{\partial\tau}+\varepsilon_\alpha\right)^{-1}.
\end{equation}
The problem is therefore mapped
on to a $(0+1)$-dimensional  problem, but, as we shall see, with a
long-ranged (temporal) interaction
due to the gaplessness of the Fermi system.

The effective action, $S_{\rm eff}$, will be expanded about the saddle
point, which is at
$\sigma(\tau)$  equal to a complex number
$\sigma_0$, independent of time, and $\lambda(\tau)=0$. Denoting the
saddle point value of the effective
action by $\widetilde{S}_{\rm eff}$, and the deviation from it by
$\delta S_{\rm eff}$, we can write
\begin{eqnarray}
\overline{F[V]-F_0[V]} &=& - \frac{1}{\beta}
\overline{\ln \frac{Z[V]}{{Z_0}[V]}}\cr
&=& \frac{1}{\beta}\left[\overline{\widetilde{S}_{\rm eff}-\ln\left(\int{\cal
D}[\cdots]e^{-\delta S_{\rm eff}}\right)}\right]
\end{eqnarray}
We shall focus on the ground state of the system; the extension to
finite, but low temperatures is
straightforward.

The saddle point value,  $\widetilde{S}_{\rm eff}$, in the limit
$\beta\to \infty$, is $\beta E_{\rm imp}^0$, where
\begin{equation}
 E_{\rm imp}^0 =\frac{N}{J_0}|\sigma_0|^2 -
\frac{N\Delta}{\pi}\left[1-\ln
\left(\frac{\Delta}{D}\right)\right]
\label{GSE}.
\end{equation}
The parameter $D$ is a high energy cutoff, and
\begin{eqnarray}
\Delta&=&\pi|\sigma_0|^2\sum_\alpha
|a_\alpha|^2\delta(\varepsilon_{\alpha})\nonumber \\
&=&\pi|\sigma_0|^2\sum_{\alpha}
|\phi_\alpha(0)|^2\delta(\varepsilon_\alpha)\label{delta}
\end{eqnarray}
If we define the local density of states at the impurity site
$\rho(\varepsilon,0)$ for a given realization
of the random potential by
\begin{equation}
\rho(\epsilon,0)=\sum_{\alpha}|\phi_{\alpha}(0)|^2\delta(\varepsilon
-\varepsilon_{\alpha}),
\end{equation}
then the extremum of Eq.~(\ref{GSE})
leads to $ E_{\rm imp}^0=-N\Delta$,
where
\begin{equation}
\Delta=D\exp\left({-\frac{1}{\rho(0,0) N |J|}}\right)\label{gap}
\end{equation}
One can easily check that the average
occupation of the $f$-level is indeed $N/2$.

The saddle point breaks the symmetry
\begin{equation}
f_M\to e^{i\theta} f_M, c_{kM}\to c_{kM},
\end{equation}
which must not be broken in the exact ground state of  this
$(0+1)$-dimensional
model for the antiferromagnetic Kondo coupling\cite{Anderson2}.

To treat the fluctuations we adapt the analysis of Witten\cite{Witten}
to include both disorder and the
long-ranged nature of the effective
$(0+1)$-dimensional model.  The insight of Witten is that we are allowed
to expand around a non-zero
vacuum expectation value of $\sigma$, attained in the $N\to\infty$
limit, but  must not break the  symmetry
$\theta\to
\theta+c$, that is,  we must not  assume a vacuum expectation value,
such as zero, for the phase of
$\sigma$.

Let us write, $\sigma(\tau)=|\sigma(\tau)|e^{i\theta(\tau)}$ and note
that it is the fluctuation of
the massless field $\theta(\tau)$ that restores the symmetry broken at
the saddle point level. Thus, to
analyze the infrared behavior, we can  set the magnitude
$|\sigma(\tau)|$ approximately to  a constant,
$|\sigma_0|$, equal to its saddle point value. Similarly, we can set
$\lambda$ to its saddle point value, which is zero. Then,
$\delta S_{\rm `eff}$ is given by
\begin{equation}
\delta S_{\rm eff}\approx
-N\mbox{Tr}\ln\left[1-|\sigma_0|^2G\left(
e^{-i\theta}G_0e^{i\theta}-G_0\right)\right]\label{deltaS}
\end{equation}
where the Green function $G_0$ can be approximated in the limit of large
imaginary time, $|\tau|\gg D^{-1}$,
by
\begin{equation}
G^0(\tau-\tau')=-\to
-\rho(0,0)\frac{P}{\tau-\tau'},\quad \beta\to \infty,\label{G0}
\end{equation}
where $P$ stands for the principal value.  We have also assumed that
$\rho(\varepsilon,0)$ is approximately independent of $\varepsilon$
around $\varepsilon=0$; below, we
shall discuss this assumption more critically. We can also calculate the
Green function
\begin{equation}
G=-\left(\frac{\partial}{\partial\tau}+|\sigma_0|^2 G^0\right)^{-1},
\end{equation}
in the long imaginary time limit. Explicitly, for $|\tau|\gg
\Delta^{-1}$,
\begin{equation}
G(\tau-\tau')
\to
-\frac{1}{\pi\Delta}\frac{P}{\tau-\tau'},\quad \beta\to \infty.
\label{G}
\end{equation}
Recalling the definition of $\Delta$ from Eq.~\ref{delta}, and the
expressions in Eqs.~\ref{G0} and \ref{G},
it is easy to see that the entire dependence on
$\Delta$ cancels from the argument of the logarithm in Eq.~\ref{deltaS},
hence the entire dependence on
disorder drops out from the fluctuation determinant at the saddle point!
The disorder averaged free energy
then simplifies enormously to
\begin{equation}
\label{eqn:fluct}
\overline{F[V]-F_0[V]}=\frac{1}{\beta}\left[
\overline{\widetilde{S}}_{\rm eff}-\ln\left(\int{\cal
D}[\cdots]e^{-\delta S_{\rm eff}}\right)\right]
\end{equation}
Ordinarily, it would have been necessary to introduce replicas
to take the disorder average of the logarithm
of the functional integral in the second term on
the right-hand-side of (\ref{eqn:fluct}).
Since the argument of the logarithm is independent
of disorder, however, replicas are unnecessary.

As in Anderson's theorem for weakly disordered
superconductors, one can argue that the local density of states at the
impurity site at the Fermi energy,
$\rho(0,0)$, cannot  differ appreciably from the density of states
averaged over the whole sample
$\overline\rho$, which, in turn, is the same as
that in the pure system. This amounts to the statement that the
distribution of the local density of
states is very narrow and centered at
the density of states of the pure
system. Thus, from Eq.~\ref{gap},
$\Delta$ is unaffected by disorder.

In fact, one can make a stronger statement, applicable
even in the localized regime, as was pointed out by Ma and
Lee\cite{Anderson1} in the context of disordered
superconductors. In the localized regime, the density of states reflects
a singular continuous spectrum with
$\delta$-function spikes, but, if  $\overline{\rho}\Delta\xi^d \gg 1$, where $d$ is the
spatial dimensionality, the theorem appears to be true. It should
break down beyond this regime, where the strong fluctuations due to disorder
will no longer allow us to assume a
smooth variation of the local density of states $\rho(\varepsilon,0)$.
Dobrosavljevi{\'c}, {\it et al.} \cite{Kotliar} have noted
that $\rho(0,0)$ is log-normally distributed at the
metal-insulator transition; such a broad distribution
can lead to the breakdown of the theorem.
These fluctuations should also induce
strong fluctuations in the amplitude of
$|\sigma(\tau)|$, frozen in the present
analysis to the saddle point
value.

It is important to note, however, that within the
stated regime of validity we have proven {\em more} than Anderson's
theorem because we have also shown that
the quantum fluctuations of the phase,
$\theta(\tau)$, is unaffected by disorder. Thus, the strong-coupling
picture of the Kondo effect, captured
correctly by the large-$N$ theory,
should be the same in the disordered
metal.

To complete our analysis it is convenient to obtain a low energy form of
the effective action $\delta S_{\rm
eff}$. To this purpose it is sufficient to expand the logarithm in
Eq.~\ref{deltaS} to get
\begin{equation}
\delta S_{\rm eff}=\frac{2N}{\pi^2}\int_{0}^{\infty}d\tau
\int_{0}^{\infty}d\tau'\frac{\sin^2\left(
\frac{\theta(\tau)-\theta(\tau')}{2}\right)}{(\tau-\tau')^2},
\label{quadratic}
\end{equation}
where we have set the upper limits of the integrals to $\infty$, as
$\beta\to \infty$ when the
ground state is approached.  This result is far more general than its
derivation suggests. This is
the only scale invariant term in $(0+1)$-dimension that one can
construct, which also respects the
compactness of the phase variables
$\theta$.  By expanding
$\sin^2((\theta(\tau)-\theta(\tau')/2)$ in Eq.~\ref{quadratic}, it
is  easy to show that
\begin{equation}
\langle \sigma^*(\tau)\sigma(0)\rangle \approx
|\sigma_0|^2\left(\frac{1}{|\tau|\Lambda}\right)^{\frac{1}{2N}}, \tau\to
\infty,
\label{corr}
\end{equation}
that is the broken symmetry is restored by the fluctuations of the
phase, but the correlation falls off very
slowly in the limit $N\to \infty$; here $\Lambda$ is a cutoff. Although
our derivation is
different, this is
a known result\cite{Read2,Coleman2}. Our analysis shows that the decay
of the correlation function is
unchanged from the pure Kondo problem. An equivalent result for the
$(1+1)$-dimensional
$SU(N)$ Thirring model was first derived and elucidated in detail  by
Witten\cite{Witten}.

As has been noted earlier\cite{Read2,Coleman2,Witten}, the ``almost
long-range order" expressed in
Eq.~\ref{corr} is sufficient to justify the conclusions drawn from the
large-$N$ expansion. Because of the
slow decay of  correlations, the electrons see the saddle point behavior
on time scales of order
$\Lambda^{-1}$, which is  bigger or of order $\Delta^{-1}$. The
elementary excitations are
expected to be fermions with a resonant
density of states not substantially different from that obtained at the
saddle point level. Therefore, it is
physically meaningful that $\Delta$ is unshifted by disorder.

Since the low-temperature
fixed point involves the formation
of a singlet bound state at the impurity, as in the
pure case, the magnetic impurity is screened. All that
is left is a unitary scatterer, which gives a universal
contribution to the resistivity.
A number of qualifying remarks should be noted before a comparison with
experiments is attempted. (1) We have  considered a situation in which the coupling $J$ itself is not a random variable.
The case in which $J$ itself is strongly random is better treated by the
theories in Refs.~\cite{Bhatt,Kotliar}. (2) We have discussed the low
temperature strong coupling regime
below the Kondo temperature and not the higher temperature
logarithmic crossover regime. (3) Our
theory is valid in the dilute magnetic impurity limit
and not in the spin glass regime. (4) Because the electron-electron
interaction is not included, its effects must be subtracted
before comparing with experiments.

The experiments which are in closest
contact to these ideas are those
of Chandrasekhar {\em et al.}\cite{Chandrasekhar}
in which the low temperature
resistivity of AuFe wires in the dilute magnetic impurity limit
was measured as a function of wire
width, temperature, and magnetic field. When the effect of
electron-electron interactions is
subtracted, the remaining Kondo contribution to the resistivity is
independent of the width of the sample
for a fixed concentration of magnetic impurities. Similar behavior is
observed for the magnetic field
dependent resistivity.
If the increase in the sheet resistance, $R_{\Box}$, as a function of the width of the films
is interpreted as  a measure of the disorder,
then these experiments can be interpreted
as evidence for the disorder independence of the
low-temperature Kondo singlet formation. The variation in 
$R_{\Box}$ was too narrow to constitute a stringent test however.

As remarked in
Ref.~\cite{Chandrasekhar}, most of the
experiments of Giordano and his
collaborators\cite{Giordano} were performed at
temperatures comparable to or larger than
the Kondo temperature, which is the regime targeted by
the theory of Martin {\em et al.}\cite{Martin}.
However, some of the data of \cite{Giordano} are in
the low-temperature regime, and they show
 a Kondo contribution to the resistivity which
decreases as the film thickness is decreased --
or, as the disorder is increased, according to the
increase in $R_{\Box}$. These results appear to contradict
our theory, but \'Ujs\'aghy, {\it et al.} \cite{Zawadowski}
have sugested that the size dependence is due to the
existence of a spin-orbit coupling-induced inert surface layer
which is proportionately more important as the
film thickness is decreased. If this interpretation
is correct, then the data of \cite{Giordano}
are consistent with those of \cite{Chandrasekhar}
and with our theory. In fact they appear to
support an even stronger statement that the
full crossover function is independent of
disorder. The experiments of DiTusa {\em
et al.}\cite{DiTusa} do not appear to be in the
dilute magnetic impurity limit. Clearly, further precise experiments will 
be very valuable.

The authors would like to thank A. Kamenev for a conversation
which inspired this work and  R. Bhatt,
A. Zawadowski,  and especially H.-Y. Kee
for discussions. S. C. acknowledges the
grant NSF-DMR-9971138. The work was also partly conducted under the
auspices of the Department
of Energy, supported (in part) by funds provided by the University of
California for the conduct
of discretionary research by Los Alamos National Laboratory.


\begin{references}
\bibitem{Anderson1}P. W. Anderson, J. Phys. Chem. Solids {\bf 11}, 26
(1959); for a
more recent perspective, see M. Ma and
P. A. Lee, Phys. Rev. B {\bf 32}, 5658 (1985).
\bibitem{Markowitz}D. Markowitz and L. P. Kadanoff, Phys. Rev. {\bf
131}, 563 (1963).
\bibitem{Martin}I. Martin, Yi Wan, and P. Phillips, Phys. Rev. Lett.
{\bf 78}, 114 (1997).
\bibitem{Coqblin}B. Coqblin and J. R. Schrieffer, Phys. Rev. {\bf 185},
847 (1969).
\bibitem{largeN}S. Chakravarty, Bull. Am. Phys. Soc. March (1982), and
unpublished.
\bibitem{Coleman}P. Coleman and N. Andrei, J. Phys. C {\bf 19}, 3211
(1986).
\bibitem{Read}N. Read and D. M. Newns, J. Phys. C {\bf 16}, 3273 (1983)
\bibitem{Read2}N. Read, J. Phys. C {\bf 18}, 2651
(1985).
\bibitem{Coleman2}P. Coleman, Phys. Rev. B {\bf 29}, 3035 (1984).
\bibitem{Anderson2}P. W. Anderson, G. Yuval and D. R. Hamann, Phys. Rev.
B {\bf 1}, 4464 (1970).
\bibitem{Witten}E. Witten, Nucl. Phys. B {\bf 145}, 110 (1978).
\bibitem{Bhatt}R. N. Bhatt and D. S. Fisher, Phys. Rev. Lett. {\bf 68},
3072 (1992).
\bibitem{Kotliar}V. Dobrosavljevi{\'c}, T. R. Kirkpatrick, and G. Kotliar,
Phys. Rev. Lett. {\bf 69}, 1113
(1992).
\bibitem{Chandrasekhar}V. Chandrasekhar {\em et al.}, Phys. Rev. Lett.
{\bf 72}, 2053 (1994).
\bibitem{Giordano}M. Blachly and N. Giordano, Phys. Rev. B {\bf 51}, 12
537 (1995), and refernces therein.
\bibitem{Zawadowski}
O. \'Ujs\'aghy and A. Zawadowski,
Phys. Rev. B {\bf 57}, 11598 (1998), and references therein.
\bibitem{DiTusa}J. F. DiTusa {\em et al.}, Phys. Rev. Lett. {\bf 68},
678 (1992).

\end{references}
\end{document}